\documentclass[12pt]{article}
\usepackage{amssymb,amsmath,amsthm,amsfonts,amscd}
\usepackage{graphicx}
\usepackage{here}

\newcommand\be{\begin{equation}}
\newcommand\ee{\end{equation}}
\newcommand\bea{\begin{eqnarray}}
\newcommand\eea{\end{eqnarray}}

\newcommand{\fatalpha}{{\bf \alpha \kern -0.44em \alpha}}
\newcommand{\fatsigma}{{\bf \sigma \kern -0.54em \sigma}}
\newcommand{\tpchi}{{\bf D \kern -0.35em D}}
\newcommand{\llambda}{{\bf \lambda \kern -0.45em \lambda}}

\bibliography{plain}
\title{\bf Entanglement of Multipartite Fermionic Coherent States for Pseudo Hermitian Hamiltonians }\vspace{20mm}
\author{ G. Najarbashi $^{a}$
 \thanks{E-mail:najarbashi@uma.ac.ir}  ,
 M.A.Fasihi $^{b,c}$
 \thanks{E-mail:fasihi@alice.math.kindai.ac.jp},
 M.Nakahara $^{b,d}$
 \thanks{E-mail:nakahara@math.kindai.ac.jp},
 F. Mirmasoudi  $^{a}$
 ,
  S. Mirzaei  $^{a}$
 \\ $^a${\small Department of Physics, University of Mohaghegh Ardabili, Ardabil, 179, Iran.} \\ $^b${\small Research Center for Quantum Computing, Interdisciplinary Graduate
School of Science and}$$\\$${\small Engineering, Kinki University, 3-4-1 Kowakae, Higashi-Osaka, Osaka 577-8502, Japan.}\\ $^c${\small Department of Physics, Azarbaijan University of Tarbiat Moallem, 53714-161, Tabriz, Iran. }\\$^d${\small Department of Physics, Kinki University, 3-4-1 Kowakae, Higashi-Osaka, Osaka 577-8502, Japan. }} \pagebreak

\vspace{20mm}
\begin{document}
\maketitle \vspace{15mm}
\newpage
\abstract{
In this paper the entanglement of multi-qubit fermionic pseudo Hermitian coherent states (FPHCS)
described by anticommutative Grassmann numbers is studied. The pseudo-Hermitian versions
of the well known maximally entangled pure states such as Bell and GHZ, W and biseparable states
are introduced through integrating over the tensor products of FPHCSs with suitable choice of
Grassmannian weight functions.
Meanwhile as an illustration, the method is applied to tensor product of 2 and 3 qubit pseudo Hermitian systems.
Then the measures of concurrence and average entropy are applied to quantify the entanglement of the pseudo two and three qubit states respectively.
}
{\bf Keywords: Pseudo Hermitian,  Entanglement, Pseudo fermionic coherent states, Pseudo Bell states,
Pseudo GHZ states, Pseudo Werner states}
\maketitle
\section{Introduction}
Quantum information theory has recently increased
its theoretical self-consistency introducing several
outstanding results.
The most important one has been the achievement that entanglement
phenomena of quantum states \cite{nielsen} have been framed
in robust theoretical schemes and verified through some
experimental tests \cite{G¨uhne,van,Carteret,Horodecki,Horodecki1,Wal,Schmid,Vidal,Bell}.
In fact entanglement is the most interesting and meanwhile strange feature of quantum physics.
The idea of entanglement starts from the apparent conflict
between the superposition principle and the nonseparability
of the related quantum states. It happens
when a state of two or more subsystems of a compound
quantum system cannot be factorized into pure local
states of the subsystems too. This is equivalent to say
that an entangled state could be used to steer a distant particle into one of a set of states, with a certain probability.
\par Furthermore the recent researches in theoretical physics and quantum optics have
revealed the importance of the coherent states. They can be used to encode quantum information
on continuous variables \cite{Llo}.
While the entanglement of the bosonic $su(2)$ and $su(1,1)$ coherent states, as the non orthogonal states which are playing an important role in the quantum cryptography and quantum information processing, has been widely investigated in the references \cite{Enk,Enk1,Fujii,wang1,wang2,wang3,wang4,wang5}, the entanglement properties of multipartite fermionic coherent states are remained as a challenging problem of quantum information theory, even from theoretical point of view \cite{Borsten,Khanna,Caste,Eshshahghader1,Eshshahghader2,Eshshahghader3}.
The fermionic coherent states are defined as the eigen-states of the annihilation operator with Grassmannian eigenvalues
\cite{majid,cabra,trifonov,Eshshahghader}. On the other hand, the last decade have witnessed a growing interest in non-Hermitian Hamiltonians with real
spectra \cite{Sch,B1,B2,B3,Can,Z,Mos1,Mos2}. Considering the results of various numerical studies, Bender and
his collaborators \cite{B1,B2} found certain examples of one-dimensional non-Hermitian Hamiltonians
that possessed real spectra. Because these Hamiltonians were invariant under PT transformations,
their spectral properties were linked with their PT-symmetry. Later Mostafazadeh
introduced the notion of pseudo-Hermiticity as an alternative possible approach for a non-
Hermitian operator to admit a real spectrum \cite{Mos1,Mos2}.
\par Recently in \cite{Eshshahghader1}, the entanglement of Grassmannian coherent states
for multi-partite n-Level Hermitian systems have been investigated considering tensor product of one mode fermionic coherent states ( e.g,  $|\theta_1\rangle|\theta_2\rangle$), defined as, $\left| {\theta} \right\rangle=\left| 0 \right\rangle- \theta \left| 1 \right\rangle$, which is presented in terms of standard basis,($\left|0 \right\rangle, \left|1 \right\rangle$) and anticommuting Grassmann numbers $ \theta_{i}\theta_{j}= - \theta_{j}\theta_{i}$. This rule is justified in the context of quantum field theory, where for example the
tensor product of two one-particle states is a two particle state and so on. Then authors found standard maximal entangled Bell, GHZ and W states by integrating over tensor product of two, three and multi-mods fermionic coherent states with proper weight functions.  The goal of this paper is extension of the presented method to the pseudo Hermitian systems.
For pseudo Hermitian systems, instead of above standard basis we deal with two set of basis $\{|\psi_{0}\rangle, |\psi_{1}\rangle\}$ and $\{|\phi_{0}\rangle,  |\phi_{1}\rangle\},$ which are the eigen-states of $H$ and $H^{\dag}$ respectively. Therefore two possible FPHCSs are
$|\theta\rangle=|\psi_{0}\rangle - \theta |\psi_{1}\rangle$ and $|\tilde{\theta}\rangle= |\phi_{0}\rangle - \theta |\phi_{1}\rangle.$
\par The paper is divided in two main parts. The first part is devoted to construction of the different families of pseudo Hermitian version of well known maximally entangled pure states such as Bell, GHZ, W and pseudo biseparable states through integrating over the tensor product of FPHCSs of two and three one qubit pseudo Hermitian system with suitable choice of Grassmannian weight functions. Then in section 2 we give a brief introduction about pseudo Hermitian quantum mechanics and in section 3 using the results of generalized Grassmaniann pseudo Hermitian coherent state \cite{Eshshahghader} we present the FPHCSs as a special case of generalized Grassmannian pseudo Hermitian coherent states for 2 level system. In section 4 we construct pseudo Hermitian version of Bell states, W and GHZ states. In the second part, section 5, we use the measures of concurrence and average entropy to quantify the entanglement of the pseudo Bell states and GHZ and Werner states respectively and discuss about the results comparing with Hermitian maximal entangled pure states. Finally conclusion is given in section $6$.
\section{Pseudo-Hermitian Hamiltonians and Biorthonormal Eigenbasis}
Intensive study of Schrodinger equation with complex potentials, but with real spectrum, was performed by different methods. The pioneer papers \cite{Sch,B1,B2,B3,Can,Z,Mos1,Mos2}, initiated investigation of $\mathrm{PT}$ symmetric systems and afterwards more general class of pseudo-Hermitian models was introduced by Mostafazadeh \cite{Mos1,Mos2}.
Following the second approach, let $H : \mathcal{H}
\rightarrow \mathcal{H}$ be a linear operator acting in a Hilbert
space $\mathcal{H}$ and $\eta : \mathcal{H} \rightarrowtail
\mathcal{H}$ be a linear Hermitian automorphism (invertible
transformation). Then the $\eta$-pseudo-Hermitian adjoint of $H$ is
defined by
\begin {equation}\label{ps1}
H^{\sharp} = \eta^{-1}H^\dag\eta.
\end{equation}
$H$ is said to be pseudo-Hermitian with respect to $\eta$ or
simply $\eta$-pseudo-Hermitian if $H^{\sharp} = H$. The eigenvalues of pseudo-Hermitian Hamiltonian
$H$ are either real or come in complex-conjugate pairs and the
following relations in nondegenerate case  hold:
\begin {equation}\label{pseudo}
    H^{\dag}=\eta H \eta^{-1}.
\end{equation}
For diagonalizable operators
$H$ with discrete spectrum, there exist a complete biorthonormal
eigenbasis $\{|\psi_{i}\rangle, |\phi_{i}\rangle\}$
such that
\begin {equation}\label{Hamiltonian}
    \begin{array}{c}
  H|\psi_{i}\rangle=E_{i}|\psi_{i}\rangle,\quad
H^{\dag}|\phi_{i}\rangle=\bar{E}_{i}|\phi_{i}\rangle, \\
  \langle\phi_{i}|\psi_{j}\rangle=\delta _{ij}, \\
  \sum_{i}|\psi_{i}\rangle\langle\phi_{i}|=\sum_{i}|\phi_{i}\rangle\langle\psi_{i}|=I. \\
\end{array}
\end{equation}
For a given pseudo-Hermitian $H$ there are infinitely many $\eta$
satisfying Eq.(\ref{pseudo}). These can however be expressed in
terms of a complete biorthonormal basis of $H$. In non degenerate case the explicit form of $\eta$ and it's inverse
satisfying Eq.(\ref{pseudo}) read
\begin {equation}\label{eta}
    \eta=\sum_{i}|\phi_{i}\rangle\langle\phi_{i}|,\qquad \eta^{-1}=\sum_{i}|\psi_{i}\rangle\langle\psi_{i}|
\end{equation}
\begin {equation}\label{phipsietamenha}
    |\phi_{i}\rangle=\eta|\psi_{i}\rangle,\qquad
    |\psi_{i}\rangle=\eta^{-1}|\phi_{i}\rangle.
\end{equation}
Through out the paper, pseudo Hamiltonian $H$ and  consequently $\eta$ and $\eta^{-1}$ assumed to be in two dimentional Hilbert space.
\section{Fermionic Pseudo-Hermitian Coherent States}
\subsection{Grassmannian variables}
The basic properties of Grassmann variables are
discussed in Refs.\cite{A,J.O,B,Glauber}
For our purpose, here, we survey the properties of this algebra which is generated
by the variables ($\theta_{1},\theta_{2}, ..., \theta_{n}$) satisfying, by definition, the following properties:
\begin {equation}\label{qdeform1}
    \begin{array}{c}
    \theta_{i}\theta_{j}= - \theta_{j}\theta_{i}\quad, \quad
i,j=1,2,...n \\
    \hspace{-3cm}\theta_{i}^2=0 .\\
  \end{array}
\end{equation}
Analogous rules also apply for the Hermitian conjugate of $\theta$,
$\theta^\dag=\bar{\theta}$, as:
\begin {equation}\label{qdeform2}
    \begin{array}{c}
    \bar{\theta}_{i}\bar{\theta}_{j}= - \bar{\theta}_{j}\bar{\theta}_{i}\quad, \quad
 i,j=1,2,...n \\
   \hspace{-3cm} \bar{\theta}_{i}^2=0 .\\
  \end{array}
\end{equation}
Any linear combination of $\theta_{i}$ with the complex coefficients is called Grassmann number. In
other words, Taylor expansion of a Grassmann function reads
$$g(\theta_{1}, \theta_{2}, ...\theta_{n}) = c_{0} + \sum_{i=1}c_{i}\theta_{i}+\sum_{i,j}{c_{i,j}\theta_{i}\theta_{j}}+...,$$
where $c_{0}, c_{i}, c_{i,j},...$ are complex numbers.
For instance, $\exp(\theta_{1}\theta_{2}) = 1 + \theta_{1}\theta_{2}$.
Integration and differentiation over complex Grassmann variables are given by Berezin's rules as :
\begin {equation}\label{brezin}
\left\{\begin{array}{c}
\hspace{-55mm}\int d\theta f(\theta)= \frac{\partial f(\theta)}{\partial \theta},\\
\int d\theta =0,~~~\int d\theta \theta=1,~~~\int d\bar{\theta}=0,~~~\int d\bar{\theta}\bar{\theta}=1, \\
\hspace{-10mm}\frac{\partial}{\partial \theta} \theta=1,~~~\frac{\partial}{\partial \theta}1=0,~~~\frac{\partial}{\partial \bar{\theta}} \bar{\theta}=1,~~~\frac{\partial}{\partial \bar{\theta}}1=0,\\
\hspace{-50mm}\frac{\partial^2}{\partial \theta ^2}=0,~~~\frac{\partial^2}{\partial \bar{\theta}^2}=0.
 \end{array}\right.
\end{equation}
To compute the integral of any
function over the Grassmann algebra the following relations are
required.
\begin {equation}\label{gG}
     \left\{\begin{array}{cc}
  \theta d \bar{\theta} = - d \bar{\theta} \theta \quad, & \bar{\theta} d \theta = -  d \theta \bar{\theta} \\
  \theta d \theta = - d \theta \theta \quad, & \bar{\theta} d \bar{\theta} = - d \bar{\theta} \bar{\theta} \\
  d \theta d \bar{\theta} = - d \bar{\theta} d\theta\quad, & \  \theta  \bar{\theta} = -  \bar{\theta} \theta. \\
\end{array}\right.
\end{equation}
\subsection{Coherent States }
Following \cite{trifonov,Eshshahghader}, One can construct the pseudo fermionic coherent states for two level pseudo hermitian Hamiltonian.
Here we outline the main results
. Considering the bi-orthonormality nature
of pseudo-Hermitian systems, we can define two pairs of annihilation and
creation operators corresponding to the bi-orthonormal eigen-states
($|\psi_{i}\rangle$,$|\phi_{i}\rangle$) respectively as
\begin {equation}\label{bsharp1}
\left\{\begin{array}{cc}
\hspace{-17mm}b:=|\psi_{0}\rangle\langle\phi_{1}|+\sqrt{{2}}\
|\psi_{1}\rangle\langle\phi_{2}|,\\
b^{\sharp}:= \eta^{-1}b^{\dag}\eta=
|\psi_{1}\rangle\langle\phi_{0}|+\sqrt{{2}} |\psi_{2}\rangle\langle\phi_{1}|,
\end{array}\right.
\end {equation}
\begin {equation}\label{bsharp2}
\left\{\begin{array}{cc} \hspace{-25mm}\tilde{b}=\eta b
\eta^{-1}=
|\phi_{1}\rangle\langle\psi_{0}|+\sqrt{{2}}\
|\phi_{2}\rangle\langle\psi_{1}|,\\
\tilde{b}^{\sharp
'}=\eta'^{-1}b^{\dag}\eta=|\phi_{0}\rangle\langle\psi_{1}|+\sqrt{{2}}\
|\phi_{1}\rangle\langle\psi_{2}|,\quad\eta'^{-1}=\eta.
\end{array}\right.
\end {equation}
Then it is possible to construct  two families of coherent states for two level pseudo Hermitian Grassmannian system in terms of
$|\psi_{k}\rangle$  and $|\phi_{k}\rangle$. The FPHCSs corresponding to
$|\psi_{k}\rangle$,  $|\phi_{k}\rangle$ denoted by
$|\theta\rangle$ and $|\tilde{\theta}\rangle$
respectively, by definition are the eigen-states of the annihilation
operators $b$ and $\tilde{b}$
\begin {equation}\label{coherent}
\left\{\begin{array}{cc}
b\ |\theta\rangle =\theta\ |\theta\rangle,\\
\tilde{b}\ |\tilde{\theta}\rangle =\theta \
|\tilde{\theta}\rangle,
\end{array}\right.
\end {equation}
and up to normalization factors are
\begin {equation}\label{coherent2}
\left\{\begin{array}{cc} |\theta\rangle=
|\psi_{0}\rangle - \theta |\psi_{1}\rangle,\\
|\tilde{\theta}\rangle= |\phi_{0}\rangle - \theta |\phi_{1}\rangle.
\end{array}\right.
\end {equation}
The explicit forms of the two families of FPHCS and characteristic
of bi-orthonormality of pseudo-Hermitian systems can be exploited
for identification of the possible integrals of $|\theta\rangle$ and
$|\tilde{\theta}\rangle$, that is $|{\theta}\rangle\langle\tilde{\theta}|$
and $|\tilde{\theta}\rangle\langle{\theta}|$, against the measure of
$d\bar{\theta}\ d\theta\  w(\theta, \bar{\theta})$, which lead to
the resolution of identity
\begin {equation}\label{resolution2}
\int d\bar{\theta}\ d\theta\  w(\theta, \bar{\theta})\
|\theta\rangle\langle\tilde{\theta}| = \int d\bar{\theta}\ d\theta\
w(\theta, \bar{\theta})\ |\tilde{\theta}\rangle\langle\theta|=I,
\end {equation}
where $w(\theta, \bar{\theta})=1+\theta\bar{\theta}.$ The Eq.(\ref{resolution2}) is called bi-over-completeness relation.
To compute weight function we require the following quantization relations between the biorthonormal
eigen-states ${|\psi_{k}\rangle},\ {|\phi_{k}\rangle}$,
$({k=0,1})$ and Grassmannian variables $\theta,
\bar{\theta}$.
\begin {equation}\label{thetapsiphi}
\left\{\begin{array}{cc}
 \theta \ |\psi_{k}\rangle = (-1)^{^{k-1}} \ |\psi_{k}\rangle\
 \theta , & \ \bar{\theta}\ \langle\psi_{k}|\ = (-1)^{^{k-1}} \ \langle\psi_{k}|\ \bar{\theta}, \\
  \theta \ \langle\psi_{k}|\ = (-1)^{^{k-1}} \
\langle\psi_{k}|\ \theta , & \ \bar{\theta} \ |\psi_{k}\rangle \
= (-1)^{^{k-1}} \ |\psi_{k}\rangle\
\bar{\theta}, \\
  \theta \ |\phi_{k}\rangle \ =(-1)^{^{k-1}} \ |\phi_{i}\rangle \
 \theta, &  \bar{\theta} \ \langle\phi_{k}| \ = (-1)^{^{k-1}} \  \langle\phi_{k}|\ \bar{\theta},
 \\
  \theta \ \langle\phi_{k}|\ = (-1)^{^{k-1}} \
\langle\phi_{k}| \ \theta, & \ \bar{\theta} \ |\phi_{k}\rangle
\ = (-1)^{^{k-1}} \ |\phi_{k}\rangle \ \bar{\theta}.
\end{array}\right.
\end {equation}
The above discussion makes it clear that neither the integral
$|\theta\rangle\langle\theta|$, nor the integral of
$|\tilde{\theta}\rangle\langle\tilde{\theta}|$, against the measure of
$d\bar{\theta}\ d\theta\  w(\theta, \bar{\theta})$ normalized:
\begin {equation}\label{resolution1}
\int d\bar{\theta}\ d\theta\  w(\theta, \bar{\theta})\
|\theta\rangle\langle\theta| \neq I, \qquad \int d\bar{\theta}\
d\theta\ w(\theta, \bar{\theta})\
|\tilde{\theta}\rangle\langle\tilde{\theta}|\neq I.
\end {equation}
One can show that the fermionic coherent states (\ref{coherent2}) remain coherent for all the times, provided that the time evolution of the initial states managed by Hamiltonian is also an eigen state of lowering operators.
\section{Maximal Pseudo Entangled States}
Suppose a fermionic system for which the particles can go  to the  $n-$mode channels. To this end, we consider tensor product of $n$ one-mode FPHCSs, each one governed by pseudo Hermitian Hamiltonians. For simplicity  we consider $n=2,3$. The case of arbitrary $n$ is straightforward.
Now we    introduce the pseudo-Hermitian  version of the well known maximally entangled pure  two  and three qubit states, such as Bell
,GHZ and W  states, \cite{Achin} respectively through integrating over the tensor product of FPHCSs  with suitable choice of Grassmannian weight functions.
\subsection{Pseudo Bell-like States}
Let us start with un-normalized pseudo Hermitian version of standard Bell states,
\begin{equation}
\begin{array}{cc}
\left| \Psi^{\pm}\right\rangle=\frac{1}{\sqrt{2}}(|01\rangle \pm |10\rangle),\\
\left| \Phi^{\pm}\right\rangle=\frac{1}{\sqrt{2}}(|00\rangle \pm |11\rangle)\\
\end{array}
\end{equation}
 that is
\begin{equation}\label{Bellstate1}
\left| {B_1}^{-}\right\rangle =\left| {\psi _0 } \right\rangle \left| {\psi _1 } \right\rangle  - \left| {\psi _1 } \right\rangle \left| {\psi _0 } \right\rangle.
\end{equation}
To achieve the above state we consider the tensor product of two one-mode FPHCSs with the same Grassmann numbers as
\begin{equation}
\left| \theta  \right\rangle \left| \theta  \right\rangle
= \left| {\psi _0 } \right\rangle \left| {\psi _0} \right\rangle  + \theta \left( {\left| {\psi _0 } \right\rangle \left| {\psi _1 } \right\rangle  - \left| {\psi _1 } \right\rangle \left| {\psi _0 } \right\rangle } \right).
\end{equation}
As mentioned already, such method is enlightened in the context of
 quantum field theory.
To get the above equation we use the explicit form of, $\left| \theta  \right\rangle$, from $Eq.(\ref{coherent2})$ . For the next step,
our task is to find the proper weight function $w(\theta)$ such that the integration over Grassmann numbers $\theta$, leads to the Eq.(\ref{Bellstate1}). To this aim let:
\begin{equation}\label{Bell}
\int {d\theta\; w(\theta)} \left| \theta  \right\rangle \left| \theta  \right\rangle  = \left| {B_1}^{-} \right\rangle,
\end{equation}
putting $w(\theta)= c_ 0 +c_ 1 \theta,$ in the above equation yields $ c_0  = 1 $~~and~~$c_1  = 0 $,
then the appropriate weight function takes the form: $w(\theta ) = 1 .$
Considering the tensor product of $\left| \theta  \right\rangle \tilde{\left|\theta \right\rangle},
\tilde{\left| { \theta } \right\rangle} \left| \theta  \right\rangle$ and $\tilde{\left|\theta \right\rangle}\tilde{\left|\theta \right\rangle}$,
,with $w(\theta)=1$, it is also possible to construct the other forms of pseudo Bell states as
\begin{equation}\label{Bellstate234}
\begin{array}{l}
\left| {B_2}^{-} \right\rangle=\int {d\theta\; } \left| \theta  \right\rangle \tilde{\left|\theta \right\rangle}  = \left| {\psi _0 } \right\rangle \left| {\varphi _1 } \right\rangle  - \left| {\psi _1 } \right\rangle \left| {\varphi _0 } \right\rangle,\\
 \end{array}
\end{equation}
So far, we concerned with tensor product of two one-mode FPHCSs with the same Grassmannian numbers, $\theta,$  and obtained the pseudo Hermitian versions of $\left| \Psi^{-}\right\rangle$. In order to establish the other pseudo Bell states
we need to consider the tensor product of FPHCSs with different Grassmann numbers, i.e.,
\begin{equation}\label{newBell}
 \left| {\theta _1 } \right\rangle \left| {\theta _2 } \right\rangle
  = \left| {\psi _0 } \right\rangle \left| {\psi _0 } \right\rangle  + \theta _2 \left| {\psi _0 } \right\rangle \left| {\psi _1 } \right\rangle  - \theta _1 \left| {\psi _1 } \right\rangle \left| {\psi _0 } \right\rangle  + \theta _1 \theta _2 \left| {\psi _1 } \right\rangle \left| {\psi _1 } \right\rangle,
\end{equation}
in this case the general form of the weight function is
$w(\theta _1, \theta _2) = c_0  + c_1 \theta _1  + c_2 \theta _2  + c_3 \theta _1 \theta _2.$
The task is to find $w(\theta _1, \theta _2)$ such that, in addition to above $\left| {B_i}^{-} \right\rangle$,
the other 3 families of pseudo Bell states are achieved. We denote these 3 families by $\left| {B_i^{+} } \right\rangle$ and $\left| B_i'^{\pm} \right\rangle$.  The results summarized in the following table.
\begin{table}[h]
\renewcommand{\arraystretch}{1}
\addtolength{\arraycolsep}{-2pt}
$$
\begin{array}{|c|c|c|c|}\hline
\rm{ state} & \rm{ FPHCS } &
\rm{weight ~ function}& \rm{psedo~ Bell~ state}
\\ \hline
  |{B_1} ^{\pm}\rangle &\left| {\theta _1 }  \right\rangle \left| {\theta _2 } \right\rangle & -(\theta _1 \pm \theta _2) & \left| {\psi _0 }  \right\rangle \left| {\psi _1 } \right\rangle  \pm \left| {\psi _1 } \right\rangle \left| {\psi _0 } \right\rangle\\
   |{B_2} ^{\pm}\rangle &\left| {\theta _1 }  \right\rangle \tilde{\left|\theta _2 \right\rangle} & -(\theta _1 \pm \theta _2) & \left| {\psi _0 } \right\rangle \left| {\phi _1 } \right\rangle  \pm \left| {\psi _1 } \right\rangle \left| {\phi _0 } \right\rangle \\
  |{B_3} ^{\pm}\rangle &\tilde{\left| {\theta _1 }  \right\rangle} \left|\theta _2 \right\rangle & -(\theta _1 \pm \theta _2) & \left| {\phi _0 } \right\rangle \left| {\psi _1 } \right\rangle  \pm \left| {\phi _1 } \right\rangle \left| {\psi _0 } \right\rangle \\
   |{B_4} ^{\pm}\rangle &\tilde{\left| {\theta _1 }  \right\rangle} \tilde{\left|\theta _2 \right\rangle}& -(\theta _1 \pm \theta _2) & \left| {\phi _0 } \right\rangle \left| {\phi _1 } \right\rangle  \pm \left| {\phi _1 } \right\rangle \left| {\phi _0 } \right\rangle \\\hline
  \hline
   |{B'_1} ^{\pm}\rangle &\left| {\theta _1 }  \right\rangle \left| {\theta _2 } \right\rangle & -(\theta _1 \theta _2 \pm 1) & \left| {\psi _0 } \right\rangle \left| {\psi _0 } \right\rangle  \pm \left| {\psi _1 } \right\rangle \left| {\psi _1 } \right\rangle\\
   |{B'_2} ^{\pm}\rangle &\left| {\theta _1 }  \right\rangle \tilde{\left|\theta _2 \right\rangle} & -(\theta _1 \theta _2 \pm 1) & \left| {\psi _0 } \right\rangle \left| {\phi _0 } \right\rangle  \pm \left| {\psi _1 } \right\rangle \left| {\phi _1 } \right\rangle \\
  |{B'_3} ^{\pm}\rangle &\tilde{\left| {\theta _1 }  \right\rangle} \left|\theta _2 \right\rangle & -(\theta _1 \theta _2 \pm 1) & \left| {\phi _0 } \right\rangle \left| {\psi _0 } \right\rangle  \pm \left| {\phi _1 } \right\rangle \left| {\psi _1 } \right\rangle \\
   |{B'_4} ^{\pm}\rangle &\tilde{\left| {\theta _1 }  \right\rangle} \tilde{\left|\theta _2 \right\rangle} & -(\theta _1 \theta _2 \pm 1) & \left| {\phi _0 } \right\rangle \left| {\phi _0 } \right\rangle  \pm \left| {\phi _1 } \right\rangle \left| {\phi _1 } \right\rangle \\
  \hline\hline
  |{B_1} ^{-}\rangle &\left| {\theta }  \right\rangle \left| {\theta } \right\rangle &1 & \left| {\psi _0 }  \right\rangle \left| {\psi _1 } \right\rangle  - \left| {\psi _1 } \right\rangle \left| {\psi _0 } \right\rangle\\
   |{B_2} ^{-}\rangle &\left| {\theta }  \right\rangle \tilde{\left|\theta \right\rangle} &1 & \left| {\psi _0 } \right\rangle \left| {\phi _1 } \right\rangle  - \left| {\psi _1 } \right\rangle \left| {\phi _0 } \right\rangle \\
  |{B_3} ^{-}\rangle &\tilde{\left| {\theta }  \right\rangle} \left|\theta \right\rangle & 1 & \left| {\phi _0 } \right\rangle \left| {\psi _1 } \right\rangle  - \left| {\phi _1 } \right\rangle \left| {\psi _0 } \right\rangle \\
   |{B_4} ^{-}\rangle &\tilde{\left| {\theta }  \right\rangle} \tilde{\left|\theta  \right\rangle}&1 & \left| {\phi _0 } \right\rangle \left| {\phi _1 } \right\rangle  - \left| {\phi _1 } \right\rangle \left| {\phi _0 } \right\rangle \\\hline
  \end{array}
$$
\caption{Unnormalized pseudo Bell states and corresponding weight functions. For example the pseudo Bell state  $\left| {\phi _0 } \right\rangle \left| {\psi _1 } \right\rangle  + \left| {\phi _1 } \right\rangle \left| {\psi _0 } \right\rangle$
  can be obtained considering tensor product $\tilde {\left| {\theta _1 } \right\rangle} {\left| { \theta _2 } \right\rangle}$ with $w(\theta _1,\theta _2)=(-\theta _1 - \theta _2)$. }\label{tab1}
\renewcommand{\arraystretch}{1}
\addtolength{\arraycolsep}{-3pt}
\end{table}
\subsection{Pseudo GHZ and W States}
In the previous subsection using possible tensor product of two one mode FPHCSs
we introduced pseudo Bell states.
Let us now proceed to construct pseudo version of the following GHZ and W states,
 \begin{equation}
\begin{array}{cc}
\left|{GHZ}^{\pm}\right\rangle=\frac{1}{\sqrt{2}}(|000\rangle \pm |111\rangle),\\
\left|W\right\rangle=\frac{1}{\sqrt{3}}(|100\rangle + |010\rangle+ |001\rangle),
\end{array}
\end{equation}
 which are used broadly in quantum information theory. To construct 3 qubit peudo GHZ, we need to consider the tensor product of 3 one mode FPHCSs, with different Grassmann numbers, where they can take following 8 forms
\begin{equation}\label{PGpossible}
\left\{\begin{array}{l}
{ \left| {\theta _1 } \right\rangle }\left| { \theta _2 } \right\rangle \left| { \theta _3 } \right\rangle,~\tilde { \left| {\theta _1 } \right\rangle }\left| { \theta _2 } \right\rangle \left| { \theta _3 } \right\rangle, ~\left| {\theta _1 } \right\rangle \tilde{\left| { \theta _2 } \right\rangle }\left| { \theta _3 } \right\rangle,\left| {\theta _1 } \right\rangle \left| { \theta _2 } \right\rangle \tilde{\left| { \theta _3 } \right\rangle}\\
\tilde { \left| {\theta _1 } \right\rangle }\tilde{\left| { \theta _2 } \right\rangle }\left| { \theta _3 } \right\rangle,~ \tilde { \left| {\theta _1 } \right\rangle }{\left| { \theta _2 } \right\rangle }\tilde{\left| { \theta _3 } \right\rangle},~\left| {\theta _1 } \right\rangle \tilde{\left| { \theta _2 } \right\rangle } \tilde{\left| { \theta _3 } \right\rangle},~\tilde { \left| {\theta _1 } \right\rangle }\tilde{\left| { \theta _2 } \right\rangle }\tilde{\left| { \theta _3 } \right\rangle}.
\end{array}\right.,
\end{equation}
As an illustration we consider the following examples:
\begin{equation}\label{PGHZ}
\begin{array}{l}
\left| {G_1^\pm } \right\rangle  =\int {d\theta_1 d\theta_2d\theta_3 w^{\pm}(\theta _1, \theta _2, \theta _3)} { \left| {\theta _1 } \right\rangle }{\left| { \theta _2 } \right\rangle }\left| { \theta _3 } \right\rangle =
 \left| {\psi _0 } \right\rangle \left| {\psi _0 } \right\rangle \left| {\psi _0 } \right\rangle  \pm \left| {\psi _1 } \right\rangle \left| {\psi _1 } \right\rangle \left| {\psi _1 } \right\rangle,\\
 \end{array}
 \end{equation}
where the weight functions are
\begin{equation}\label{weightPG}
w^{\pm}(\theta _1,  \theta _2,  \theta _3 )=\theta _3 \theta _2 \theta _1  \pm  1.
\end{equation}
One can easily check that the appropriate weight function for each of the $\left| {G_i^\pm } \right\rangle,~~i=1,...,8$ is the same and equals to Eq.(\ref{weightPG}). The results for unnormalized pseudo GHZ states summarized in the the following table
\begin{table}[h]
\renewcommand{\arraystretch}{1}
\addtolength{\arraycolsep}{-2pt}
$$
\begin{array}{|c|c|c|c|}\hline
\rm{ state} & \rm{ FPHCS } &
\rm{weight ~ function}& \rm{pseudo~ GHZ ~ state}
\\ \hline
  |{G_1 }^{\pm}\rangle &\left| {\theta _1 }  \right\rangle \left| {\theta _2 } \right\rangle\left| {\theta _3 } \right\rangle & \theta _3 \theta _2 \theta _1  \pm  1& \left| {\psi _0 } \right\rangle \left| {\psi _0 } \right\rangle \left| {\psi _0 } \right\rangle  \pm \left| {\psi _1 } \right\rangle\left| {\psi _1 } \right\rangle \left| {\psi _1 } \right\rangle \\
   |{G_2}^{\pm} \rangle &\tilde{\left| {\theta _1 }  \right\rangle} \left|\theta _2 \right\rangle\left| {\theta _3 } \right\rangle &  \theta _3 \theta _2 \theta _1  \pm  1& \left| {\varphi _0 } \right\rangle \left| {\psi _0 } \right\rangle \left| {\psi _0 } \right\rangle  \pm \left| {\varphi _1 } \right\rangle \left| {\psi _1 } \right\rangle \left| {\psi _1 } \right\rangle  \\
  |{G_3 }^{\pm}\rangle &\left| {\theta _1 }  \right\rangle \tilde{\left| {\theta _2 } \right\rangle}\left| {\theta _3 } \right\rangle &  \theta _3 \theta _2 \theta _1  \pm  1 & \left| {\psi _0 } \right\rangle \left| {\varphi _0 } \right\rangle \left| {\psi _0 } \right\rangle  \pm \left| {\psi _1 } \right\rangle \left| {\varphi _1 } \right\rangle \left| {\psi _1 } \right\rangle   \\
   |{G_4 }^{\pm}\rangle &\left| {\theta _1 }  \right\rangle \left| {\theta _2 } \right\rangle\tilde{\left| {\theta _3 } \right\rangle}&  \theta _3 \theta _2 \theta _1  \pm  1 & \left| {\psi _0 } \right\rangle \left| {\psi _0 } \right\rangle \left| {\varphi _0 } \right\rangle  \pm \left| {\psi _1 } \right\rangle \left| {\psi _1 } \right\rangle \left| {\varphi _1 } \right\rangle  \\
   |{G_5 }^{\pm}\rangle &\tilde{\left| {\theta _1 }  \right\rangle} \tilde{\left|\theta _2 \right\rangle}\left| {\theta _3 } \right\rangle & \theta _3 \theta _2 \theta _1  \pm  1 & \left| {\varphi _0 } \right\rangle \left| {\varphi _0 } \right\rangle \left| {\psi _0 } \right\rangle  \pm \left| {\varphi _1 } \right\rangle \left| {\varphi _1 } \right\rangle \left| {\psi _1 } \right\rangle  \\
   |{G_6 }^{\pm}\rangle &\tilde{\left| {\theta _1 }  \right\rangle} \left|\theta _2 \right\rangle\tilde{\left| {\theta _3 } \right\rangle} & \theta _3 \theta _2 \theta _1  \pm  1 & \left| {\varphi _0 } \right\rangle \left| {\psi _0 } \right\rangle \left| {\varphi _0 } \right\rangle  \pm \left| {\varphi _1 } \right\rangle \left| {\psi _1 } \right\rangle \left| {\varphi _1 } \right\rangle   \\
  |{G_7 }^{\pm}\rangle &\left| {\theta _1 }  \right\rangle \tilde{\left|\theta _2 \right\rangle}\tilde{\left|\theta _3 \right\rangle} &  \theta _3 \theta _2 \theta _1  \pm  1& \left| {\psi _0 } \right\rangle \left| {\varphi _0 } \right\rangle \left| {\varphi _0 } \right\rangle  \pm \left| {\psi _0 } \right\rangle \left| {\varphi _1 } \right\rangle \left| {\varphi _0 } \right\rangle \\
   {|G_8 }^{\pm}\rangle &\tilde{\left| {\theta _1 }  \right\rangle} \tilde{\left|\theta _2 \right\rangle}\tilde{\left|\theta _3 \right\rangle} & \theta _3 \theta _2 \theta _1  \pm  1 & \left| {\varphi _0 } \right\rangle \left| {\varphi _0 } \right\rangle \left| {\varphi _0 } \right\rangle  \pm \left| {\varphi _1 } \right\rangle \left| {\varphi _1 } \right\rangle \left| {\varphi _1 } \right\rangle  \\
    \hline
  \end{array}
$$
\caption{Unnormalized pseudo GHZ states and corresponding weight functions . }\label{tab1}
\renewcommand{\arraystretch}{1}
\addtolength{\arraycolsep}{-3pt}
\end{table}
To construct the pseudo W states, one can use either the tensor product of FPHCSs with 3 different or the same Grassmann numbers. In the following we introduce one example for each categories, denoted by $\mathcal{W}$ and $\mathcal{W'}$ respectively.
For tensor product of the FPHCSs with different Grassmann numbers we have
\begin{equation}\label{PW}
\begin{array}{l}
 \left| \mathcal{W}_1 \right\rangle =\int {d\theta_1 } {d\theta_2 }{d\theta_3 }\;w_1(\theta_1, \theta_2, \theta_3)\left| \theta_1  \right\rangle {\left| { \theta_2 } \right\rangle} {\left| { \theta_3 } \right\rangle }\\
 \hspace{14mm} =\left| {\psi _0 } \right\rangle \left| {\psi _0 } \right\rangle \left| {\psi _1 } \right\rangle + \left| {\psi _0 } \right\rangle \left| {\psi _1 } \right\rangle \left| {\psi _0 } \right\rangle + \left| {\psi _1 } \right\rangle \left| {\psi _0 } \right\rangle \left| {\psi _0 } \right\rangle
  \end{array}
 \end{equation}
where
\begin{equation}\label{weightw}
w(\theta_1, \theta_2, \theta_3)=\theta_1 \theta_2 + \theta_1 \theta_3 +  \theta_2 \theta_3.
\end{equation}
Similarly for the same Grassmann numbers we get
\begin{equation}\label{PW'}
\begin{array}{l}
 \left| \mathcal{W'}_1 \right\rangle =\int {d\theta } \;w(\theta){\left| \theta  \right\rangle} {\left| { \theta } \right\rangle} {\left| { \theta } \right\rangle }\\
  \hspace{14mm} =-\left| {\psi _0 } \right\rangle \left| {\psi _0 } \right\rangle \left| {\psi _1 } \right\rangle  + \left| {\psi _0 } \right\rangle \left| {\psi _1 } \right\rangle \left| {\psi _0 } \right\rangle  -\left| {\psi _1 } \right\rangle \left| {\psi _0 } \right\rangle \left| {\psi _0 } \right\rangle,
  \end{array}
 \end{equation}
 where the proper weight function is:
\begin{equation}\label{weightwp}
w'(\theta)=1.
\end{equation}
As it clears in table $III$, for a given tensor product of three different one mode FPHCSs, e.g, $\left| {\theta _1 }  \right\rangle \left| {\theta _2 } \right\rangle\left| {\theta _3 } \right\rangle$ depending on
the selection of weight function, there are eight pseudo W states.
 We emphasized that, although we construct the category $\mathcal{W'}$ in terms of FPHCSs with the same Grassmann numbers, one may also tempt to obtain the same result with the different Grassmann numbers which in turns yield the weight function
$w=-\theta_1 \theta_2 + \theta_1 \theta_3 -  \theta_2 \theta_3.$
\begin{table}[h]
\renewcommand{\arraystretch}{1}
\addtolength{\arraycolsep}{-2pt}
$$
\begin{array}{|c|c|c|c|}\hline
\rm{ state} & \rm{ FPHCS } &
\rm{weight ~ function}& \rm{pseudo~ W ~ state}
\\ \hline
  |{W_1 }^{(i)}\rangle &\left| {\theta _1 }  \right\rangle \left| {\theta _2 } \right\rangle\left| {\theta _3 } \right\rangle & \pm \theta _1 \theta _2\pm\theta _1 \theta _3\pm \theta _2 \theta _3 & \pm\left| {\psi _0 } \right\rangle \left| {\psi _0 } \right\rangle \left| {\psi _1 } \right\rangle  \pm \left| {\psi _0 } \right\rangle \left| {\psi _1 } \right\rangle \left| {\psi _0 } \right\rangle  \pm\emph{}\left| {\psi _1 } \right\rangle \left| {\psi _0 } \right\rangle \left| {\psi _0 } \right\rangle\\
   |{W_2 }^{(i)}\rangle &\tilde{\left| {\theta _1 }  \right\rangle} \left|\theta _2 \right\rangle\left| {\theta _3 } \right\rangle &  \pm \theta _1 \theta _2\pm\theta _1 \theta _3\pm\theta _2 \theta _3 & \pm\left| {\varphi _0 } \right\rangle \left| {\psi _0 } \right\rangle \left| {\psi _1 } \right\rangle  \pm \left| {\varphi _0 } \right\rangle \left| {\psi _1 } \right\rangle \left| {\psi _0 } \right\rangle  \pm\left| {\varphi _1 } \right\rangle \left| {\psi _0 } \right\rangle \left| {\psi _0 } \right\rangle \\
  |{W_3 }^{(i)}\rangle &\left| {\theta _1 }  \right\rangle \tilde{\left| {\theta _2 } \right\rangle}\left| {\theta _3 } \right\rangle &  \pm \theta _1 \theta _2\pm\theta _1 \theta _3\pm\theta _2 \theta _3 & \pm\left| {\psi _0 } \right\rangle \left| {\varphi _0 } \right\rangle \left| {\psi _1 } \right\rangle  \pm \left| {\psi _0 } \right\rangle \left| {\varphi _1 } \right\rangle \left| {\psi _0 } \right\rangle  \pm\left| {\psi _1 } \right\rangle \left| {\varphi _0 } \right\rangle \left| {\psi _0 } \right\rangle \\
   |{W_4 }^{(i)}\rangle &\left| {\theta _1 }  \right\rangle \left| {\theta _2 } \right\rangle\tilde{\left| {\theta _3 } \right\rangle}&  \pm \theta _1 \theta _2\pm\theta _1 \theta _3\pm\theta _2 \theta _3 & \pm\left| {\psi _0 } \right\rangle \left| {\psi _0 } \right\rangle \left| {\varphi _1 } \right\rangle  \pm \left| {\psi _0 } \right\rangle \left| {\psi _1 } \right\rangle \left| {\varphi _0 } \right\rangle  \pm\left| {\psi _1 } \right\rangle \left| {\psi _0 } \right\rangle \left| {\varphi _0 } \right\rangle \\
   |{W_5 }^{(i)}\rangle &\tilde{\left| {\theta _1 }  \right\rangle} \tilde{\left|\theta _2 \right\rangle}\left| {\theta _3 } \right\rangle & \pm \theta _1 \theta _2\pm\theta _1 \theta _3\pm\theta _2 \theta _3 & \pm\left| {\varphi _0 } \right\rangle \left| {\varphi _0 } \right\rangle \left| {\psi _1 } \right\rangle  \pm \left| {\varphi _0 } \right\rangle \left| {\varphi _1 } \right\rangle \left| {\psi _0 } \right\rangle  \pm\left| {\varphi _1 } \right\rangle \left| {\varphi _0 } \right\rangle \left| {\psi _0 } \right\rangle\\
   |{W_6 }^{(i)}\rangle &\tilde{\left| {\theta _1 }  \right\rangle} \left|\theta _2 \right\rangle\tilde{\left| {\theta _3 } \right\rangle} & \pm \theta _1 \theta _2\pm\theta _1 \theta _3\pm\theta _2 \theta _3 & \pm\left| {\varphi _0 } \right\rangle \left| {\psi _0 } \right\rangle \left| {\varphi _1 } \right\rangle  \pm \left| {\varphi _0 } \right\rangle \left| {\psi _1 } \right\rangle \left| {\varphi _0 } \right\rangle  \pm\left| {\varphi _1 } \right\rangle \left| {\psi _0 } \right\rangle \left| {\varphi _0 } \right\rangle \\
  |{W_7 }^{(i)}\rangle &\left| {\theta _1 }  \right\rangle \tilde{\left|\theta _2 \right\rangle}\tilde{\left|\theta _3 \right\rangle} &  \pm \theta _1 \theta _2\pm\theta _1 \theta _3\pm\theta _2 \theta _3 & \pm\left| {\psi _0 } \right\rangle \left| {\varphi _0 } \right\rangle \left| {\varphi _1 } \right\rangle  \pm \left| {\psi _0 } \right\rangle \left| {\varphi _1 } \right\rangle \left| {\varphi _0 } \right\rangle  \pm\left| {\psi _1 } \right\rangle \left| {\varphi _0 } \right\rangle \left| {\varphi _0 } \right\rangle\\
   |{W_8 }^{(i)}\rangle &\tilde{\left| {\theta _1 }  \right\rangle} \tilde{\left|\theta _2 \right\rangle}\tilde{\left|\theta _3 \right\rangle} & \pm \theta _1 \theta _2\pm\theta _1 \theta _3\pm\theta _2 \theta _3 & \pm\left| {\varphi _0 } \right\rangle \left| {\varphi _0 } \right\rangle \left| {\varphi _1 } \right\rangle  \pm \left| {\varphi _0 } \right\rangle \left| {\varphi _1 } \right\rangle \left| {\varphi _0 } \right\rangle  \pm\left| {\varphi _1 } \right\rangle \left| {\varphi _0 } \right\rangle \left| {\varphi _0 } \right\rangle\\
  \hline\hline
  |{W'_1 }\rangle &\left| {\theta }  \right\rangle \left| {\theta } \right\rangle\left| {\theta } \right\rangle & 1 & -\left| {\psi _0 } \right\rangle \left| {\psi _0 } \right\rangle \left| {\psi _1 } \right\rangle + \left| {\psi _0 } \right\rangle \left| {\psi _1 } \right\rangle \left| {\psi _0 } \right\rangle -\left| {\psi _1 } \right\rangle \left| {\psi _0 } \right\rangle \left| {\psi _0 } \right\rangle\\
   |{W'_2 }\rangle &\tilde{\left| {\theta }  \right\rangle} \left|\theta \right\rangle\left| {\theta } \right\rangle &  1 & -\left| {\varphi _0 } \right\rangle \left| {\psi _0 } \right\rangle \left| {\psi _1 } \right\rangle  + \left| {\varphi _0 } \right\rangle \left| {\psi _1 } \right\rangle \left| {\psi _0 } \right\rangle -\left| {\varphi _1 } \right\rangle \left| {\psi _0 } \right\rangle \left| {\psi _0 } \right\rangle \\
  |{W'_3 }\rangle &\left| {\theta }  \right\rangle \tilde{\left| {\theta } \right\rangle}\left| {\theta } \right\rangle &  1 & -\left| {\psi _0 } \right\rangle \left| {\varphi _0 } \right\rangle \left| {\psi _1 } \right\rangle  + \left| {\psi _0 } \right\rangle \left| {\varphi _1 } \right\rangle \left| {\psi _0 } \right\rangle  -\left| {\psi _1 } \right\rangle \left| {\varphi _0 } \right\rangle \left| {\psi _0 } \right\rangle \\
   |{W'_4 }\rangle &\left| {\theta }  \right\rangle \left| {\theta } \right\rangle\tilde{\left| {\theta } \right\rangle}& 1 & -\left| {\psi _0 } \right\rangle \left| {\psi _0 } \right\rangle \left| {\varphi _1 } \right\rangle  + \left| {\psi _0 } \right\rangle \left| {\psi _1 } \right\rangle \left| {\varphi _0 } \right\rangle  - \left| {\psi _1 } \right\rangle \left| {\psi _0 } \right\rangle \left| {\varphi _0 } \right\rangle \\
   |{W'_5 }\rangle &\tilde{\left| {\theta }  \right\rangle} \tilde{\left|\theta \right\rangle}\left| {\theta } \right\rangle &1 & -\left| {\varphi _0 } \right\rangle \left| {\varphi _0 } \right\rangle \left| {\psi _1 } \right\rangle  + \left| {\varphi _0 } \right\rangle \left| {\varphi _1 } \right\rangle \left| {\psi _0 } \right\rangle  -\left| {\varphi _1 } \right\rangle \left| {\varphi _0 } \right\rangle \left| {\psi _0 } \right\rangle\\
   |{W'_6 }\rangle &\tilde{\left| {\theta }  \right\rangle} \left|\theta \right\rangle\tilde{\left| {\theta } \right\rangle} & 1 & -\left| {\varphi _0 } \right\rangle \left| {\psi _0 } \right\rangle \left| {\varphi _1 } \right\rangle  + \left| {\varphi _0 } \right\rangle \left| {\psi _1 } \right\rangle \left| {\varphi _0 } \right\rangle  -\left| {\varphi _1 } \right\rangle \left| {\psi _0 } \right\rangle \left| {\varphi _0 } \right\rangle \\
  |{W'_7 }\rangle &\left| {\theta }  \right\rangle \tilde{\left|\theta \right\rangle}\tilde{\left|\theta \right\rangle} & 1 & -\left| {\psi _0 } \right\rangle \left| {\varphi _0 } \right\rangle \left| {\varphi _1 } \right\rangle  + \left| {\psi _0 } \right\rangle \left| {\varphi _1 } \right\rangle \left| {\varphi _0 } \right\rangle - \left| {\psi _1 } \right\rangle \left| {\varphi _0 } \right\rangle \left| {\varphi _0 } \right\rangle\\
   |{W'_8 }\rangle &\tilde{\left| {\theta }  \right\rangle} \tilde{\left|\theta  \right\rangle}\tilde{\left|\theta  \right\rangle} & 1 & -\left| {\varphi _0 } \right\rangle \left| {\varphi _0 } \right\rangle \left| {\varphi _1 } \right\rangle  + \left| {\varphi _0 } \right\rangle \left| {\varphi _1 } \right\rangle \left| {\varphi _0 } \right\rangle - \left| {\varphi _1 } \right\rangle \left| {\varphi _0 } \right\rangle \left| {\varphi _0 } \right\rangle\\
 \hline
 \end{array}
$$
\caption{Unnormalized pseudo $W$ states and corresponding weight functions. The upper index $(i)$ refers to set of symbols \{(+,+,+),~ (+,+,-),...(-,-,-)\}, addressing to the set of weight functions \{$ \theta _1 \theta _2 + \theta _1 \theta _3 + \theta _2 \theta _3,~ \theta _1 \theta _2 + \theta _1 \theta _3 - \theta _2 \theta _3,~ -\theta _1 \theta _2 - \theta _1 \theta _3 - \theta _2 \theta _3$ \} respectively.}\label{tab1}
\renewcommand{\arraystretch}{1}
\addtolength{\arraycolsep}{-3pt}
\end{table}
\subsection{Pseudo Biseparable States}
Here we use FPHCSs to construct pseudo biseparabile states. Depending on how one considers bi-partition
for a given state, there exists an entanglement in their subsystems partially. For example
if a pure state $\left| ABC \right\rangle$ involves the three subsystems A,B and C, the partition {A} may be
separable while {B,C} are entangled. For illustration, let us consider the following examples
\begin{equation}
\left\{ \begin{array}{l}
\int{d\theta _1 d\theta _2 d\theta_3(\theta_ 1 \theta_ 2\pm \theta_ 1  \theta_ 3)\left| {\theta _1 } \right\rangle \left| {\theta _2 } \right\rangle \left| {\theta _3 } \right\rangle}={\left| {\psi _0 } \right\rangle}_{(1)} \otimes {\left| {B _1 }^\pm \right\rangle }_{(2,3)},\\
\int{d\theta _1 d\theta _2 d\theta_3(\theta_ 3 \theta_ 2\theta_ 1 \mp  \theta_ 1)\left| {\theta _1 } \right\rangle \left| {\theta _2 } \right\rangle \left| {\theta _3 } \right\rangle}={\left| {\psi _0 } \right\rangle}_{(1)} \otimes {\left| {B' _1 }^\pm \right\rangle }_{(2,3)},\\
\int{d\theta _1 d\theta _2 d\theta_3 (\theta_ 1 \theta_ 2 \mp \theta_ 3 \theta_ 2)\left| {\theta _1 } \right\rangle \left| {\theta _2 } \right\rangle \left| {\theta _3 } \right\rangle}={\left| {\psi _0 } \right\rangle}_{(2)} \otimes {\left| {B _1 }^\pm \right\rangle }_{(1,3)},\\
\end{array} \right.
\end{equation}
where the first two examples show that the partition $(2,3)$ is pseudo Bell state and is separable with respect to partition $1$.  The same statement holds for partitions $(1,3)$ and $2$ in the last example.
The above examples make it clear that, we could find different pseudo biseparable states just by considering the integration over tensor product  $\left| {\theta _1 } \right\rangle \left| {\theta _2 } \right\rangle \left| {\theta _3 } \right\rangle$ using different weight functions.
However one should note that the family $W'$ does not lead to any pseudo biseparable states.
\section{Entanglement of Multipartite pseudo Hermitian states }
So far, what we have achieved is the constructing of the pseudo Hermitian version of Bell, GHZ and W states. In what follows, we will study the entanglement of pseudo Bell using the measure of concurrence and pseudo GHZ and W states by means of average entropy.
To this end let us consider, the following two level PT symmetric \cite{BenPRL} or pseudo Hermitian Hamiltonians
\begin{equation}\label{PH}
H_i = \left( {\begin{array}{*{20}c}
   {r_ie^{i\beta_i } } & s_i  \\
   t_i & {r_ie^{ - i\beta_i } }  \\
\end{array}} \right),~~~i=1,2,3
\end{equation}
where index $i$ stands for $i^{th}$ system. We assume that the systems, reside in four and eight dimensional Hilbert space, are governed by $ H_1 \otimes H_2 $ and $ H_1\otimes H_2\otimes H_3 $  respectively.
The bi-orthonormal eigen-sates of $H_i$, and $H_i^\dag$ are
\begin{equation}\label{explicit psi and phi}
\left\{\begin{array}{l}
 \left| {\psi _0 } \right\rangle^{(i)}  = \frac{1}{\sqrt {2 \cos \alpha_i }}
 ( \begin{array}{l}
 e^{\frac{{i\alpha_i }}{2} },\;
e^{ - \frac{{i\alpha_i }}{2} }\end{array} )^T \\
\left| {\psi _1 } \right\rangle^{(i)}  = \frac{1}{\sqrt{ 2 \cos \alpha_i }}( \begin{array}{l}
 e^{\frac{{-i\alpha_i }}{2} },
  - e^{ \frac{{i\alpha_i }}{2}} \end{array})^T
 \end{array}\right.,~~~
 \left\{\begin{array}{l}
 \left| {\varphi _0 } \right\rangle^{(i)}  = \frac{1}{\sqrt{ 2 \cos \alpha_i }}( \begin{array}{l}
 e^{\frac{{-i\alpha_i }}{2} },\; e^{\frac{{i\alpha_i }}{2} } \end{array} )^T\\
 \left| {\varphi _1 } \right\rangle^{(i)}  = \frac{1}{\sqrt{ 2 \cos \alpha_i }}(\begin{array}{l}
 e^{\frac{{i\alpha_i }}{2} },\;\;
  - e^{ - \frac{{i\alpha_i }}{2} }
 \end{array})^T
\end{array}\right.,
\end{equation}\\
where $\sin \alpha_i=\frac{r_i}{{\sqrt {s_it_i} }}\sin \beta_i,$  and
$T$ denotes the transpose. In the next subsection we first consider the pseudo Bell states.
\subsection{Entanglement of pseudo Bell states}
It is well known that the entanglement of a two-qubit state $|\psi\rangle$ can be expressed as a function of
concurrence \cite{Hill,Wootters}
\begin{equation}\label{concurrence}
\mathcal{C}(|\psi\rangle)\equiv|\langle\psi|\sigma_{y}\otimes\sigma_{y}|\psi^{*}\rangle|
\end{equation}
where $\sigma_{y}$ is the y component of the Pauli matrices and $|\psi^{*}\rangle$ is the complex conjugate of $|\psi\rangle$.
Since concurrence itself can also be considered as a measure of entanglement \cite{Wootters}, in the following we use it to quantify the entanglement of pseudo Bell states.
After normalizing all pseudo Bell states mentioned in table $I$ and recalling the explicit forms of $(\left| {\psi_{k}  } \right\rangle^{(i)} $ and $ \left| {\varphi_{k}  } \right\rangle^{(i)},\;k=0,1) $, from Eq.(\ref{explicit psi and phi}), the corresponding concurrences take the following forms
\begin{equation}\label{CondiffpraB}
 \left\{\begin{array}{l}
 \mathcal{C}( \left| B_{1} ^{-}\right\rangle)=\mathcal{C}( \left| B_{4}^{-} \right\rangle)=|\frac{\cos \alpha_1 \cos \alpha_2}{1-\sin \alpha_1\sin \alpha_2}|, \\
 \mathcal{C}( \left| B_{2} ^{-}\right\rangle)=\mathcal{C}( \left| B_{3}^{-} \right\rangle)=|\frac{\cos \alpha_1 \cos \alpha_2}{1+\sin \alpha_1\sin \alpha_2}|,\\
 \end{array}\right.
 \end{equation}
 where we focused on the third part of table $I$. Similar discussion can be made for other pseudo Bell states.
 So the concurrence of $\left| B_{j}^{-}\right\rangle$s is a periodic function with respect to the parameters $\alpha_1$ and $\alpha_2$ with the period $T=\pi$, that is
$ \mathcal{C}(\alpha_1, \alpha_2)=\mathcal{C}(\alpha_1+m \pi,\alpha_2+m \pi)$, where m belongs to integer numbers.
The above equations show that for both cases  $\mathcal{C}_{max}=1$ and $\mathcal{C}_{min}=0$ appear in $\alpha_{1}=\alpha_{2}= m\pi$ and
$\alpha_{1}=\alpha_{1}=(2m+1)\frac{\pi}{2}$ respectively.
For the special case $\alpha_1=\alpha_2=\alpha$ Eq.(\ref{CondiffpraB}) reads
 \begin{equation}\label{entanglement}
 \left\{\begin{array}{l}\emph{}
 \mathcal{C}( \left| B_{1} ^{-}\right\rangle)=\mathcal{C}( \left| B_{4} ^{-}\right\rangle)=1, \\
 \mathcal{C}( \left| B_{2} ^{-}\right\rangle)=\mathcal{C}( \left| B_{3} ^{-}\right\rangle)=\frac{\cos^2(\alpha)}{1+\sin^2(\alpha)}.\\
 \end{array}\right.
 \end{equation}
It should be no surprise that we obtain $\mathcal{C}=1$ for $\left| B_{1}^{-} \right\rangle$ and $\left| B_{4}^{-} \right\rangle\emph{}$, independent of parameter $\alpha$, since for $\alpha_1=\alpha_2=\alpha$ these states reduce to standard Bell state $|\Psi^{-}\rangle$ up to the total phase $e^{-i\pi}$,
 \begin{equation}\label{gammaB}
\left\{\begin{array}{l}
  \left| B_{1} \right\rangle^{-} = \frac{ {\left| {\psi _0 } \right\rangle \left| {\psi _1 } \right\rangle  - \left| {\psi _1 } \right\rangle \left| {\psi _0 } \right\rangle }}{\|\left| B_{1}^{-} \right\rangle\|}  =- \frac{1}{{\sqrt{2}}}({\left| {01} \right\rangle  - \left| {10} \right\rangle })  ,\\
 \left| B_{4} \right\rangle^{-} = \frac{ {\left| {\varphi _0 } \right\rangle \left| {\varphi _1 } \right\rangle  - \left| {\varphi _1 } \right\rangle \left| {\varphi _0 } \right\rangle } }{\|\left| B_{4}^{-} \right\rangle\|}= - \frac{1}{{\sqrt{2}}}({\left| {01} \right\rangle  - \left| {10} \right\rangle }).
 \end{array}\right.
 \end{equation}
In contrast, the concurrence of $ \left| B_{2}^{-} \right\rangle$ and $ \left|B_{3}^{-} \right\rangle$ varying as a function of $\alpha$ is depicted in Fig.1.
\begin{figure}[H]
 \begin{center}
  \includegraphics[width=8cm]{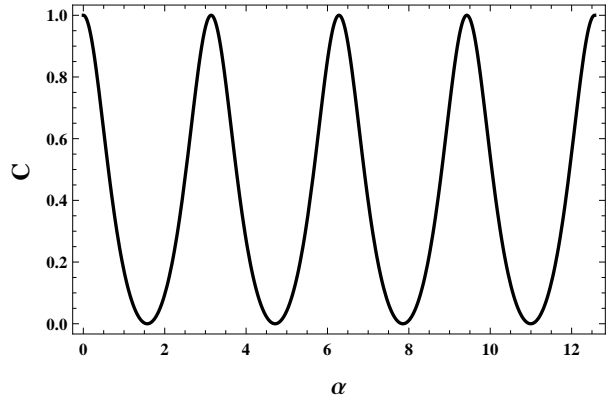}
\caption{Concurrence of $\left| B_{2}^{-}\right\rangle$ and $\left| B_{3}^{-}\right\rangle$ as function of the parameter $\alpha$.
}
  \label{graph:C1}
 \end{center}
\end{figure}
Simple calculation reveals that for $\alpha_1 =\alpha_2=\alpha$  the following pseudo Bell states reduce to standard Bell states.
\begin{equation}
\begin{array}{l}
\left| {B'_2}^{-} \right\rangle = \left| {B'_3}^{-} \right\rangle= \left| \Psi^{+} \right\rangle,\\
\left| {B'_1}^{+} \right\rangle = \left| {B'_4}^{+} \right\rangle =\left| \Phi^{+} \right\rangle,\\
\left| {B_2}^{+} \right\rangle = \left| {B_3}^{+} \right\rangle =\left| \Phi^{-} \right\rangle.
 \end{array}
\end{equation}
We consider more special cases  which is interesting in dipole interaction decay as follows.
{\bf{Case a~:~}}if $st = r^2 \sin ^2 \beta$ then $\mathcal{C}(\left| B_{2}^{-} \right\rangle)=\mathcal{C}(\left| B_{3}^{-} \right\rangle)=0$\\
{\bf{Case b~:~}}if $r=\frac{\delta}{2}$,~~ $\beta= -\frac{\pi}{2}$ ,~~$t=s,$ then the Hamiltonian 32 reduce to
\begin{equation}\label{concurrence1}
H_{1,2} = \frac{1}{2}\left( {\begin{array}{*{20}c}
   { - i\delta } & 2s  \\
   2s & {i\delta }  \\
\end{array}} \right).
\end{equation}
This Hamiltonian arises in interacting two level atom with an electromagnetic field
where the real constants $\delta$ is the decay rate for the upper and lower levels and the quantity $s$ characterizes the radiation-atom interaction matrix element between the levels described in interaction picture with rotating wave approximation \cite{trifonov,lam,car}.
In this case the concurrence in terms of $s$ and $\delta$ is $\mathcal{C}( \left| B_{2}^{-} \right\rangle)=\mathcal{C}( \left| B_{3}^{-} \right\rangle)=|\frac{4s^2-\delta^2}{4s^2+\delta^2}|$.
Since $\sin\alpha=-\frac{\delta}{2s}$, then $4s^2-\delta^2\geq 0$ which guarantees the nonnegativity of concurrence.
Fig.~\ref{graph:C2} shows concurrence of $\left| B_{2 }^{-} \right\rangle$ and $\left| B_{3}^{-} \right\rangle$ for the intervals:  $ 1 \leq s \leq 2$ and $-2 \leq \delta \leq 2 $.
\begin{figure}[H]
 \begin{center}
  \includegraphics[width=8cm]{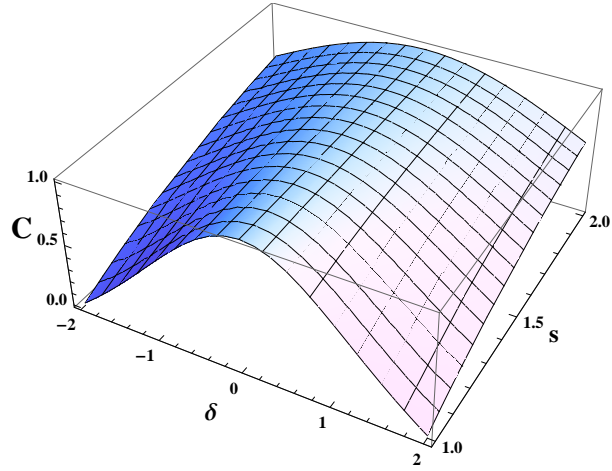}
 \caption{Concurrence of $\left| B_{2}^{-}\right\rangle$ and $\left| B_{3}^{-}\right\rangle$ in terms of the parameters $\delta$ and $s$, as it seen for the points $(\delta=0,s)$ concurrence of these states are equal to one and they are maximally entangled.
}
  \label{graph:C2}
 \end{center}
\end{figure}
\subsection{Entanglement of pseudo GHZ and W states}
\begin{figure}[H]
 \begin{center}
  \includegraphics[scale=1]{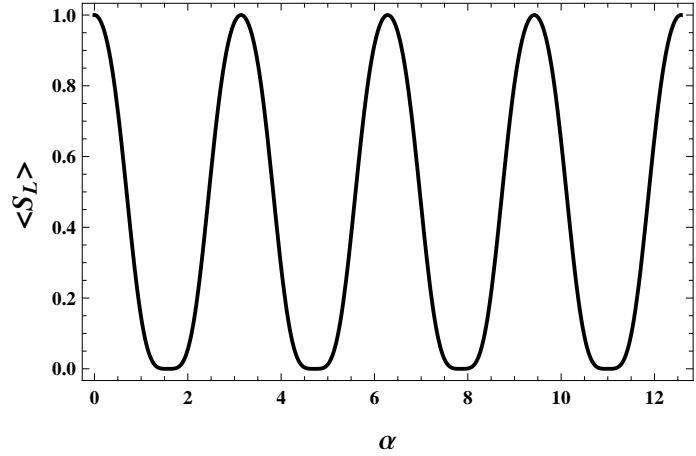}
  \caption{The average entropy of all pseudo GHZ states, verses  parameter $\alpha$.
}
  \label{graph:SLPG}
 \end{center}
\end{figure}

\begin{figure}[H]
 \begin{center}
  \includegraphics[scale=1]{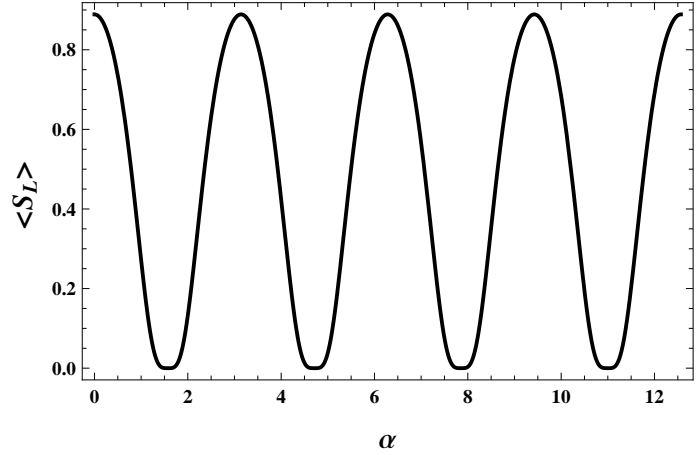}
  \caption{The average entropy of $\left| \mathcal{W}_7 \right\rangle$, as a function of $\alpha$. The lower bound $0$ and upper bound $\frac{8}{9}$  correspond to separable and maximal entangled pseudo $\left| \mathcal{W}_7 \right\rangle$ states respectively.
}
  \label{graph:SLPW}
 \end{center}
\end{figure}

\begin{figure}[H]
 \begin{center}
  \includegraphics[scale=1]{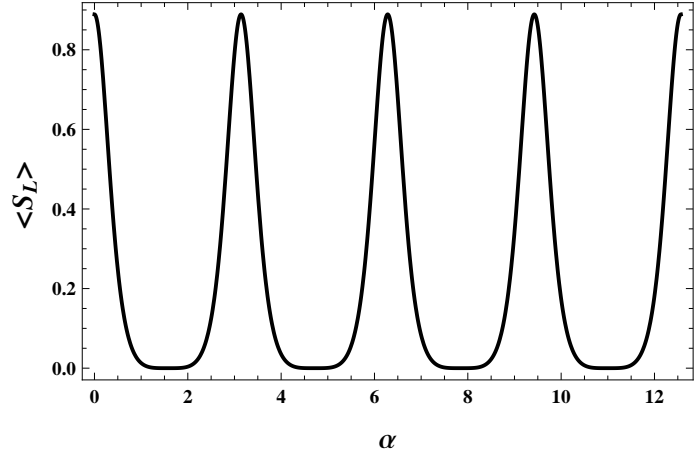}
  \caption{Average entropy of $|W_6\rangle$ as a function of $\alpha$.
}
\label{graph:SLPWp}
 \end{center}
\end{figure}

In the previous subsection we studied the entanglement of pseudo Bell states using the measure of concurrence.
For the next step we are interested to quantify the entanglement of pseudo GHZ and W states.
The well behavior measure that we shall consider is the average entropy $\langle S_{L} \rangle$,
\begin{equation}\label{average Entropy}
\left\langle {S_L } \right\rangle  = \left( \begin{array}{l}
 N \\
 n \\
 \end{array} \right)^{-1}\sum\limits_{A_n } {S_L^{(A_n ;B_{N - n} )} },
\end{equation}
which is define via linear entropy $S_{L}$\cite{Rajag},
as
\begin{equation}\label{Linear Entropy}
S_L^{(A_n ;B_{N - n} )}  = \frac{d}{{d - 1}}(1 - Tr_{A_n } \left[ {\rho _{A_n } } \right]^2 ),~~~~~\rho _{A_n }  = Tr_{B_{N - n} } [\rho ],
\end{equation}
where, $d = \min \{ 2^n ,2^{N - n} \}$, is the dimension of the reduced density matrix $\rho _{A_n }$.
It should be recall that although the linear entropy and von Neumann entropy \cite{Plenio} are similar measures of the mixedness of a state, the linear entropy is easier to calculate because it does not require the diagonalization of the density matrix.
The linear entropy can range between zero, corresponding to a completely pure state, and, $1$ corresponding to a completely mixed state
. Based on the measure of average entropy, as examples, we shall investigated the entanglement of the normalized $\left|{G_1^+}\right\rangle$,  $\left|{\mathcal{W}_7}^{(+,+,+)}\right\rangle$ and $\left|{\mathcal{W}_6}^{(-,+,-)}\right\rangle$ (for simplicity denoted by $W_7$ and $W_6$ respectively)  as
\begin{equation}\label{PGHZN}
\begin{array}{l}
\left| {G_1^+ } \right\rangle  =\frac{\left\{
 \left| {\varphi _0 } \right\rangle \left| {\varphi _0 } \right\rangle \left| {\varphi _0 } \right\rangle  \pm \left| {\varphi _1 } \right\rangle \left| {\varphi _1 } \right\rangle \left| {\varphi _1 } \right\rangle\right\}}{\parallel\left| G_{1}^+ \right\rangle\parallel}
\\
   \left| \mathcal{W}_7 \right\rangle= \frac{\left| {\psi _0 } \right\rangle \left| {\varphi _0 } \right\rangle \left| {\varphi _1 } \right\rangle + \left| {\psi _0 } \right\rangle \left| {\varphi _1 } \right\rangle \left| {\varphi _0 } \right\rangle + \left| {\psi _1 } \right\rangle \left| {\varphi _0 } \right\rangle \left| {\varphi _0 } \right\rangle}{{\parallel\left| W_{7} \right\rangle\parallel}}\\
  \left| \mathcal{W}_6 \right\rangle= \frac{-\left| {\varphi _0 } \right\rangle \left| {\psi _0 } \right\rangle \left| {\varphi _1 } \right\rangle  + \left| {\varphi _0 } \right\rangle \left| {\psi _1 } \right\rangle \left| {\varphi _0 } \right\rangle  -\left| {\varphi _1 } \right\rangle \left| {\psi _0 } \right\rangle \left| {\varphi _0 } \right\rangle}{{{\parallel\left| W_{6} \right\rangle\parallel}}}
  \end{array}
 \end{equation}
Regarding the definition of the Eq.(\ref{average Entropy}) the average entropy of the normalized $\left| {G_1^+ } \right\rangle$ is given as:
\begin{equation}\label{PGaveragegeneral}
\left\langle {S_L } \right\rangle_{_{ {(G_1^+)} }}=\frac{1}{6}\left(5+\cos 2\alpha_2 - 2\sin^2 \alpha_1(1+\cos^2 \alpha_3\sin^2 \alpha_2)+( \cos 2\alpha_1 \cos2\alpha_2-3)\sin^2 \alpha_3\right)
\end{equation}
Straightforward calculations reveal that,
the average entropy of all of the pseudo GHZ states are the same and equal to Eq.(\ref{PGaveragegeneral}).
As before, let us consider the quantum states with $\alpha_1=\alpha_2=\alpha_3=\alpha$, which yields
\begin{equation}\label{PGaverage}
 \left\langle {S_L } \right\rangle_{_{({G_1^+ })}}=\frac{1}{2}\cos ^4\alpha (3-\cos 2 \alpha). \\
 \end{equation}
Fig.~\ref{graph:SLPG} shows the average entropy of the $\left|G_1^+ \right\rangle$, in terms of the
parameter $\alpha$.
The maximum and minimum value of average entropy for pseudo GHZ states occur at the points $\alpha=k \pi$ and $\alpha=(2k+1)\frac{\pi}{2}$ respectively.
As another example, we shall consider the normalized $\left| \mathcal{W}_7 \right\rangle$,
where its average entropy for the cases of different and identical $\alpha_{i}$ respectively are
\begin{equation}
\hspace{-100mm}\left\langle {S_L } \right\rangle_{_{(\mathcal{W}_7 )}} =
\end{equation}
$$\frac{2(\cos 2 \alpha _1+\cos 2 \alpha _2+2)\cos 2 \alpha _3+\cos 2 (\alpha _1-\alpha _2)+\cos 2 (\alpha _1+\alpha _2)+4 \cos 2 \alpha _1+4 \cos 2 \alpha _2+6}{3 (2 \sin \alpha _2 \sin \alpha _3-2 \sin \alpha _1 (\sin \alpha _2+\sin \alpha _3)+3)^2}
$$
\begin{equation}
\left\langle {S_L } \right\rangle_{_{( \mathcal{W}_7 )}}  = \frac{8 \cos ^4 \alpha }{(\cos {2 \alpha} + 2)^2}.
\end{equation}
Fig.~\ref{graph:SLPW} shows that the average entropy of $\left| \mathcal{W}_7 \right\rangle$, for identical
case is bounded bye the following values
$$0 \leq \left\langle {S_L(\alpha) } \right\rangle_{_{( \mathcal{W}_7 )}}\leq \frac{8}{9},$$
where the upper and lower bounds appear at $\alpha=k \pi$ and $\alpha_{k}= (2k+1)\frac{\pi}{2}$ respectively.
Finally as the last example let us study the average entropy of the normalized $\left| \mathcal{W}_6 \right\rangle$. To this end, considering the
Eq.(\ref{average Entropy}) we deduce the following expression for $\left\langle {S_L } \right\rangle_{_{( \mathcal{W}_6 )}}$ \emph{}
\begin{equation}
\hspace{-100mm}\left\langle {S_L } \right\rangle_{_{(\mathcal{W}_6 )}} =\end{equation}
$$\frac{2(\cos 2 \alpha _1+\cos 2 \alpha _2+2)\cos 2 \alpha _3+\cos 2 (\alpha _1-\alpha _2)+\cos 2 (\alpha _1+\alpha _2)+4 \cos 2 \alpha _1+4 \cos 2 \alpha _2+6}{3 (2 \sin \alpha _2 \sin \alpha _3+2 \sin \alpha _1 (\sin \alpha _2+\sin \alpha _3)+3)^2}
$$
It is easy to check that for the case of $\alpha_1=\alpha_2=\alpha_3=\alpha$ the above equation reduce to
\begin{equation}
\left\langle {S_L } \right\rangle_{_{( \mathcal{W}_6 )}}  =  \frac{8 \cos ^4 \alpha }{9 (\cos 2 \alpha - 2)^2}.
\end{equation}
Fig.~\ref{graph:SLPWp}, shows the $\left\langle {S_L } \right\rangle_{_{(\mathcal{W}_6 )}}$ in terms of the parameter $\alpha$
and like the previous cases, the maximum value of the average entropy of the $\left| \mathcal{W}_6 \right\rangle$ is exactly the same as that of the entangled states described in standard Hermitian Hamiltonian.
The method presented can also be extend to mulitipartide $n$ level systems.
\section{Conclusion}
In conclusion we have constructed the pseudo-Hermitian version
of the well known maximally entangled pure states such as Bell GHZ, W, and biseparable states, by
integrating over tensor product of one-mode FPHCSs and  using suitable Grassmannian weight function.
Meanwhile to clarify the issue, we explicitly consider the bi-orthonormal eigen-states of pseudo Hermitian Hamiltonian
which appears in interacting two level atom with an electromagnetic field.
In order to quantify the entanglement of aforementioned pseudo states, we used concurrence measure and average linear entropy for two qubit (pseudo Bell states) and 3 qubit (pseudo GHZ and W) respectively.
It is found that for $\alpha_1=\alpha_2=\alpha$ pseudo Bell states $\left| {B_1}^{-} \right\rangle$ and $\left| {B_4}^{-} \right\rangle$ up to the total phase $e^{-i\pi}$, are the same as standard Bell state $\left| \Psi^- \right\rangle$. Similarly $\left| {B'_2}^{-} \right\rangle$ and $\left| {B'_3}^{-}\right\rangle$ are the same as $\left| \Psi^- \right\rangle$ and $\left| {B'_1}^{+} \right\rangle$ and $\left| {B'_4}^{+}\right\rangle$ are equal to $\left| \Phi^+ \right\rangle$ and $\left| {B_2}^{+} \right\rangle$ and $\left| {B_3}^{+}\right\rangle$ reduce to $\left| \Phi^- \right\rangle$.

\end{document}